\documentclass[manuscript]{aastex}

\usepackage{amssymb}
\usepackage{longtable}
\usepackage{array}
\usepackage{txfonts}

\usepackage{multirow}
\usepackage{rotating}
\usepackage{lscape}
\usepackage{amsmath}

\slugcomment{Not to appear in Nonlearned J., 45.}

\shorttitle{Force-Freeness of Solar Photosphere Magnetic Field}
\shortauthors{Liu S.}

\begin{document}


\title{A Research on the Force-Freeness of Photosphere Magnetic Field}

%% WITH SOLAR DYNAMICS OBSERVATORY/ATMOSPHERIC IMAGING ASSEMBLY

\author{S. Liu,  \altaffilmark{1} J. Hao, \altaffilmark{1}}
\email{lius@nao.cas.cn}


\altaffiltext{1}{Key Laboratory of Solar Activity, National
Astronomical Observatories, Chinese Academy of Science, Beijing
100012, China}

\begin{abstract}
In this paper, the statistical studies on the force-freeness of photosphere magnetic
fields are given. The studies are based on the vector magnetic fields observed by Solar Optical Telescope/Spectro-Polarimeter (SOT/SP) on
board Hinode.
Three parameters ($F_{x}/ F_{p}$, $F_{y}/ F_{p}$, and $F_{z}/ F_{p}$) are introduced
to investigate the force-freeness of active regions photosphere magnetic field.
Various thresholds and reductions of original resolutions are selected to calculated parameters.
As for the resolutions, the reductions of original resolution data by a factor of 2, 4 and 8 are applied.
While for thresholds, they are calculated from individual active region with the average of 75/77/57 G for $B_{x}$/$B_{y}$/$B_{z}$ for all original data.
When the resolution are reduced by 2, 4 and 8, the corresponding averages are 72/75/55, 66/68/49 and 57/59/41 G for $B_{x}$/$B_{y}$/$B_{z}$, respectively.
It is found that the forces indicated by $F_{x}/ F_{p}$ and $F_{y}/ F_{p}$ increase as thresholds/resolutions increase/decrease.
While for $F_{z}/ F_{p}$ the trends become more complex. For low threshold, when the resolution decrease the trends of $F_{z}/ F_{p}$
are similar as those of $F_{x}/ F_{p}$ and $F_{y}/ F_{p}$, while for high threshold the trends of $F_{z}/ F_{p}$ are different from those of $F_{x}/ F_{p}$ and $F_{y}/ F_{p}$. For original resolution, the forces indicated by $F_{y}/ F_{p}$ increase as thresholds increase,
while for others resolution they decrease as the thresholds increase contrarily.
%while for others resolution (reduce original data by a factor of 2, 4 and 8) they decrease as the thresholds increase contrarily.

\end{abstract}




\section{INTRODUCTION}
Magnetic fields filling the whole active regions are related directly to solar active eruption events such as filament eruptions, jets,
flares and coronal mass ejections. Hence, the knowledge of magnetic fields is necessary to understand the nature of solar activities.
Due to the whole space of active regions are filled with magnetic fields, the studies of active regions magnetic fields
should include the information of 3 dimensions magnetic fields of active regions.
Up to present, magnetic fields in the photosphere can be measured with high accuracy and
reliability. Magnetic field measurements in chromosphere and corona are only available for a few cases, due to its difficulties from both observations and physical theories \citep{gar94, lin04}.
As a result, the magnetic field extrapolations work as alternative and effective methods
to study the spatial magnetic fields in chromosphere and corona.

The conventional and classical field extrapolations are force-free extrapolations, that assume
magnetic fields above the photosphere are free of Lorentz forces \citep{aly89}.
This assumption can be considered to be appropriate for coronal environments, where are filled with
the low-$\beta$ plasma (here, $\beta$ is defined as the ratio between plasma and magnetic pressure).
Then, the corona magnetic fields can be reconstructed through force-free extrapolation basing on force-free model and using photosphere magnetic fields
as boundary conditions. Although there are lots of field extrapolations applications to study spatial magnetic
fields of active regions \citep{wu90, mic94, ama97, sak81, yan00, whe00, wie04, son06, he08, liu11a, liu11b, wie12}, there exist an inconsistency unavoidably that how the numerical solutions satisfying force-free match the photosphere magnetic fields that should be not force-free completely.
There should be a fact that the force-free extent of the photosphere magnetic fields
may affect the reliability of field extrapolations at some extent.
Recently, some researchers propose various methods to resolve this inconsistency \citep{wie06,jia13} as far as possible.
Additionally, the itself magnetic characteristics including force-free extent of magnetic fields in the photosphere are worth to study and reveal.

Recently, there are some studies about the force-free extent of magnetic fields in the photosphere are presented,
however more accurate conclusions are worth to deliberate still.
The force-freeness of photosphere magnetic field is firstly studied by \citet{mat95} basing on magnetograms observed by Mees Solar Observatory.
The changes of net Lorentz force with the height from the photosphere to low chromosphere are studied.
The conclusions given are that magnetic fields deviate from force-free in the photosphere, while magnetic field
become force-free about 400 km above surface of the photosphere. \citet{moo02} concluded that magnetic field of these active regions do not deviate from force-free too far, their studies basing on magnetograms observed by the Haleakala Stokes Polarimeter of Mees Solar Observatory.
The similar results as \citet{moo02} are given by \citet{tiw12} using high resolution magnetic field
obtained from Solar Optical Telescope/Spectro-Polarimeter (SOT/SP) aboard Hinode satellite.
However, \citet{liu13} gave a statistical study on force-freeness of magnetic field in the Photosphere
using 925 magnetograms observed by the Solar Magnetic Field Telescope located at Huairou Solar Observing Station,
the results show that the extents of unforce-free of photosphere magnetic field are higher than those drew by \citet{moo02, tiw12}.
In this paper following the above studies , the problems about the force-freeness of photosphere magnetic fields
are intensively researched through a relative large data sample with high resolution obtained from SP/Hinode.

The paper are organized as follows. The observation data and data processing are described in Section 2. The
results obtained are shown in Section 3. Finally, in Section 4, the brief conclusions and summaries are
given.

\section{OBSERVATIONS DATA}

In this paper observations data obtained by the Spectro-polarimeter (SP) aboard Hinode \citep{kos07}
are used.
For SP magnetic field observations, there are two magnetically sensitive Fe lines of 630.15 and 630.25 are employed,
When SP observing, it work at the center of those two lines or nearby continuum. SP through the slit scan to combine and generate
Stokes $IQUV$ spectral images. The magnetic and thermodynamic parameters contained in observations are inverted from full Stokes lines profiles basing on Milne-Eddington (ME) atmospheric model.
The vector magnetograms used in this paper are downloaded from CSAC site (http://www.csac.hao.ucar.edu/).
After the inversion, the normal level2 data contains 36 magnetic and thermodynamic parameters, of which the field strength B, the inclination angle $\gamma$ and the azimuth angle $\phi$ are used to create magnetograms.
The longitudinal magnetic field component
can be obtained with the formula $Bcos(\gamma)$, while the transverse field can be calculated from expression
$Bsin(\gamma)$. The acute angle method \citep{wan94, wan97,mat06} is employed to resolve transverse field $180^{\circ}$ ambiguity.

The data sample used in this work are observed from November 2006 to June 2012,
during this period about 600 active regions appeared on the sun,
but a part of active regions were not observed by SP/Hinode.
In this study, the active regions with the longitude and latitude less than 40 degree are chosen.
Additionally, for multi-observations that taken on the same active region,
the magnetograms that is closest to the center of disk are selected.
%At last, it gives a total number of 139 active regions to constitute data sample.
The data sample is extend from \cite{hao11} with the same selection criteria.
The field of view for each magnetogram keeps original one that observed by SP/Hinode,
and the resolution is 0.16/0.32 arcsec/pixel for the normal/fast modes observed magnetograms.
To avoid the projection effects, the vector magnetic fields are transformed to local heliographic coordinates,
then the magnetic components ($B_{x}$,$B_{y}$ and $B_{z}$) are generated for this study.

\section{RESULTS}

Following the previous studies \citep{mat95,moo02, tiw12,liu13}, three parameters of $F_{x}/ F_{p}$, $F_{y}/ F_{p}$, and $F_{z}/ F_{p}$
expressed as follows are used to study the force-free extent of photosphere magnetic fields:
\begin{equation}
\label{fxyzp01} F_{x} = -\dfrac{1}{4\pi}\int B_{x}B_{z}dxdy,
\end{equation}
\begin{equation}
\label{fxyzp02}F_{y} = -\dfrac{1}{4\pi}\int B_{y}B_{z}dxdy,
\end{equation}
\begin{equation}
\label{fxyzp03}F_{z} = -\dfrac{1}{8\pi}\int (B_{z}^{2}-B_{x}^{2}-B_{y}^{2})dxdy.
\end{equation}
\begin{equation}
\label{fxyzp1} F_{p} = \dfrac{1}{8\pi}\int
(B_{z}^{2}+B_{x}^{2}+B_{y}^{2})dxdy,
\end{equation}

\begin{equation}
\label{flux} Flux_{Devi.} = \int (Bz dxdy) / (\int |Bz dxdy|),
\end{equation}
Here, the assumption is that the magnetic field filling in half space above the photosphere decrease very fast when z increases, it will vanish when z goes infinite. Equations (\ref{fxyzp01})-(\ref{fxyzp03}) due to the
balance between net Lorentz force in the infinite half-space (z$>$0)
and Maxwell stress integrated on surface (z=0).
While $F_{p}$ indicates the magnitude of total Lorentz force.
\citet{mol67} also illustrated some basic integral relations about the criteria for the force-free character of active region magnetic field.
Flux balance of active regions is important for the force-free conditions (1-3),
so in this work the requirement $Flux_{Devi.}$ (Eq.\ref{flux}) less than 0.2 is added.
\textbf{At last 112 active regions are selected as data sample to be investigated.}

\citet{low85} suggest that if all three Lorentz force components are much less than $F_{p}$, then it can give a force-free field.
So, the magnitude of $F_{x}/ F_{p}$, $F_{y}/ F_{p}$, and $F_{z}/ F_{p}$ can role as a criterion to measure the force-free extent
of magnetic field at the boundary of the photosphere z = 0. \citet{mat95} suggested that if $F_{z}/ F_{p}$ is equal to or less than 0.1 ($F_{z}/ F_{p} \le 0.1$)
then the magnetic field of photosphere can be regarded as force-free completely.
\citet{moo02} analysed 12 magnetograms, of which the mean and median values of $F_{z}/ F_{p}$ are -0.17 and -0.13, respectively,
and concluded that the magnetic field of photosphere is not so far from force-free.
\citet{tiw12} analysed high resolution vector magnetograms of 19 active regions observed by SOT/SP,
of which the  mean and median values of $F_{z}/ F_{p}$ are -0.23 and -0.22, respectively, and it is suggest that
the magnetic field of photosphere do not deviate force-free case too much.

\begin{figure}

%%%I:\SP\readsav_plotzhif-sandian-sigma0123
   \centerline{\includegraphics[width=1\textwidth,clip=]{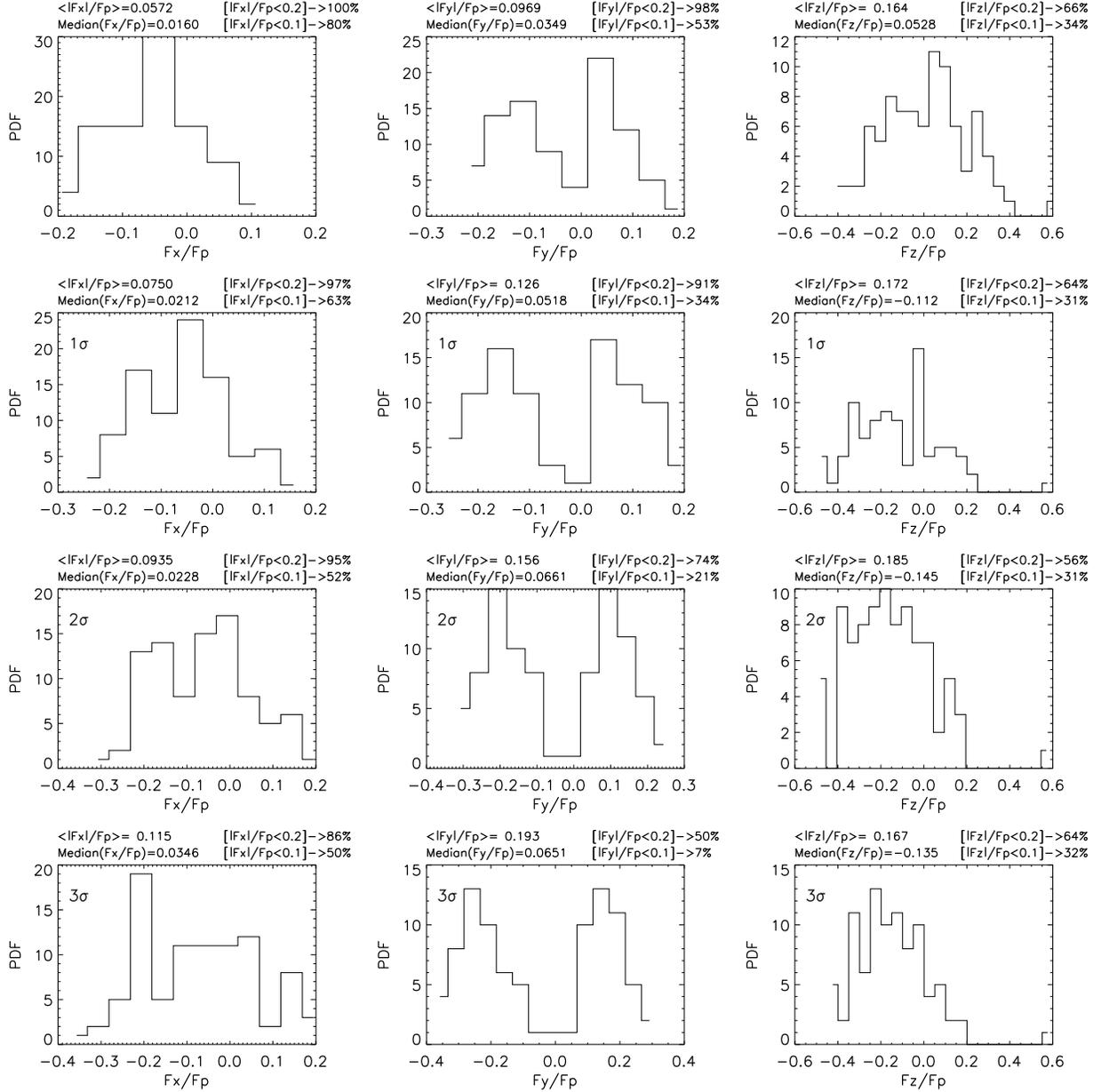}}

   \caption{PDF of $F_{x}/F_{p}$, $F_{y}/F_{p}$ and $F_{z}/F_{p}$ for the all
   selected magnetograms using some certain thresholds in calculations (in the first row the threshold is 0,
   in the second, third and fourth rows the thresholds are 1$\sigma$, 2$\sigma$ and 3$\sigma$ deviations of data noise).
   Here, the mean values of absolute $F_{x}/F_{p}$, $F_{y}/F_{p}$ and
   $F_{z}/F_{p}$ are plotted and indicated by $<|F_{x}|/F_{p}>$,$<|F_{y}|/F_{p}>$, $<|F_{z}|/F_{p}>$, respectively.
   the median of $F_{x}/F_{p}$, $F_{y}/F_{p}$ and
   $F_{z}/F_{p}$ are plotted and indicated by $Median(F_{x}/F_{p})$, $Median(F_{y}/F_{p})$, $Median(F_{z}/F_{p})$, respectively.
   Additionally, the percentages of $F_{x}/F_{p}$, $F_{y}/F_{p}$ and $F_{z}/F_{p}$ less than 0.1 and 0.2 are given correspondingly.
   } \label{Fig1}
\end{figure}

\begin{figure}

%%%
   \centerline{\includegraphics[width=1\textwidth,clip=]{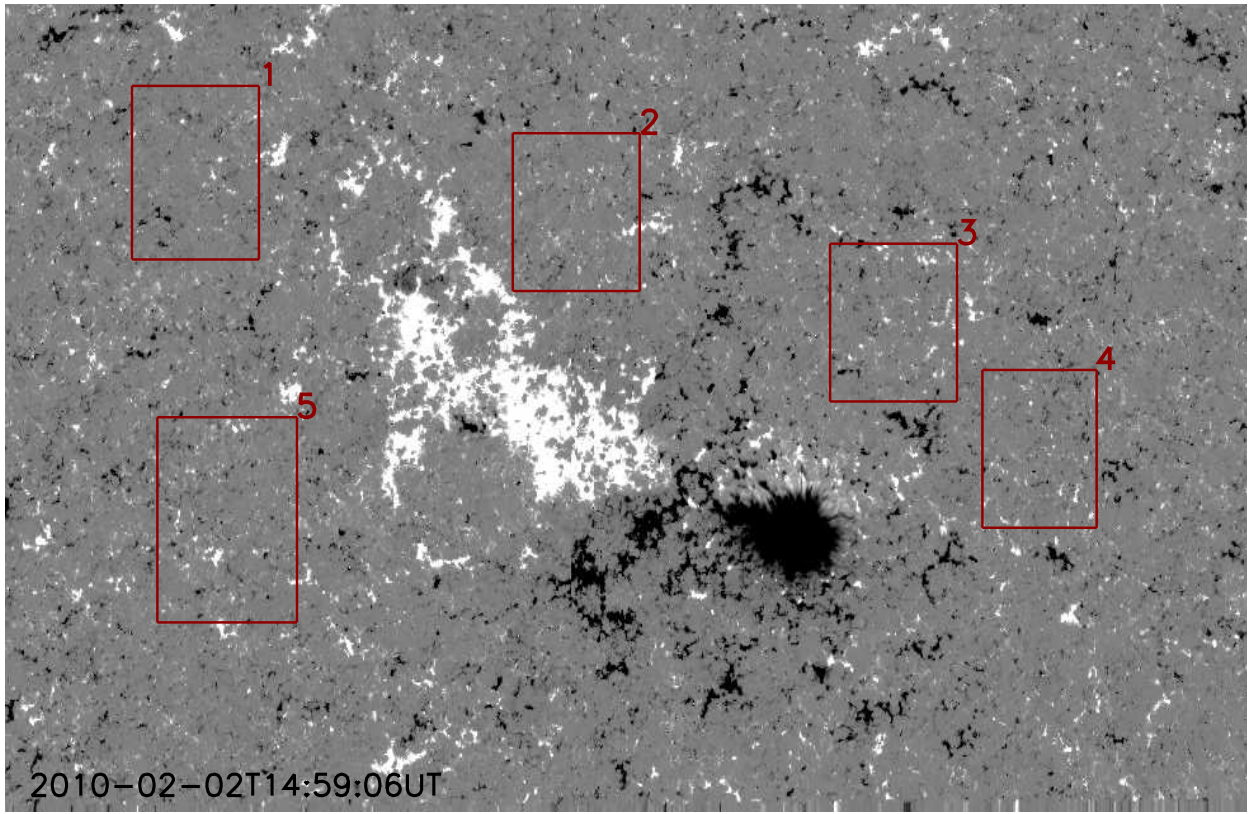}}

   \caption{An example of active region image used to show the selections of quiet region to obtain $\sigma$, the quiet regions are indicated by red rectangles.
   } \label{Fig1-add-1}
\end{figure}

Figure \ref{Fig1} shows the probability density function (PDF) of $F_{x}/ F_{p}$, $F_{y}/ F_{p}$ and $F_{z}/ F_{p}$
for the selected active regions magnetograms. To avoid the effects of data noise,
here magnetic component $B_{x}$, $B_{y}$ and $B_{z}$ values greater
than some certain threshold are calculated and presented, respectively. These thresholds are determined by the different levels of data noise,
the 1$\sigma$, 2$\sigma$ and 3$\sigma$ deviations of each magnetic components are used as different thresholds.
For each magnetogram, $\sigma$ deviation is calculated using quiet regions in individual active region,
and quiet regions are around but not closed to the main sunspot.Figure \ref{Fig1-add-1} gives an example of active region
to show how the quiet regions are selected, firstly, the individual $\sigma$ is calculated for each quiet region labeled by 1, 2, 3 et al.,
then the average of $\sigma$s
for all selected quiet regions is regarded as threshold used in this active region. For different active region the number of quiet regions maybe
different, which ranges from 3 to 6 basing on distributions of active region magnetic field.
Only pixels with $B_{x}$, $B_{y}$ and $B_{z}$ simultaneously greater than 1$\sigma$ (2$\sigma$, 3$\sigma$) deviations are calculated
to get the values of $F_{x}/ F_{p}$, $F_{y}/ F_{p}$ and $F_{z}/ F_{p}$.
In figure \ref{Fig1}, the first row shows the PDF distributions of $F_{x}/ F_{p}$, $F_{y}/ F_{p}$ and $F_{z}/ F_{p}$
that are computed from all pixels of magnetic components of $B_{x}$, $B_{y}$ and $B_{z}$, it means there is no threshold used in the calculations.
The second, third and fourth rows show the PDF distributions of of $F_{x}/ F_{p}$, $F_{y}/ F_{p}$ and $F_{z}/ F_{p}$, which are calculated using
1$\sigma$, 2$\sigma$ and 3$\sigma$ deviations as thresholds for each magnetic components, respectively.
In each plot, the median of those parameters of $F_{x}/ F_{p}$, $F_{y}/ F_{p}$ and $F_{z}/ F_{p}$, the average of absolute value ($<|F_{i}|/F_{p}>, i=x,y,z$) of those parameters are shown, additionally, the percentages of those parameters for two cases, which are the corresponding
parameters ($F_{x}/ F_{p}$, $F_{y}/ F_{p}$ and $F_{z}/ F_{p}$) are less than 0.1 and 0.2, are given to get a relative quantitative results.
It is found that $F_{x}/ F_{p}$ and $F_{y}/ F_{p}$ increase as the thresholds increase evidently, while the changes of $F_{z}/ F_{p}$ are very faint.
For $F_{x}/ F_{p}$, the percentage of active region magnetic field regarded as force-free sufficiently ($F_{x}/ F_{p}<0.1$) decrease from 80\% (threshold=0) to 50\% (threshold=3$\sigma$); For $F_{y}/ F_{p}$ the percentage decrease from 53\% (threshold=0) to 7\% (threshold=3$\sigma$); For $F_{z}/ F_{p}$ dissimilarly, this percentage almost do not change.
If it is assumed that the magnetic field can be regarded as not far from force-free model, when $F_{x}/ F_{p}$, $F_{y}/ F_{p}$ and $F_{z}/ F_{p}$ less than 0.2. Then, for $F_{x}/ F_{p}$ and $F_{y}/ F_{p}$, the percentage of active region magnetic field regarded as not far from force-free decrease from 100\%/98\% (threshold=0) to 86\%/50\% (threshold=3$\sigma$), while for $F_{z}/ F_{p}$ the corresponding percentage changes faintly.

\begin{figure}

%%%I:\document-1\force-free-SP\read-all-dde-sm-all
   \centerline{\includegraphics[width=1\textwidth,clip=]{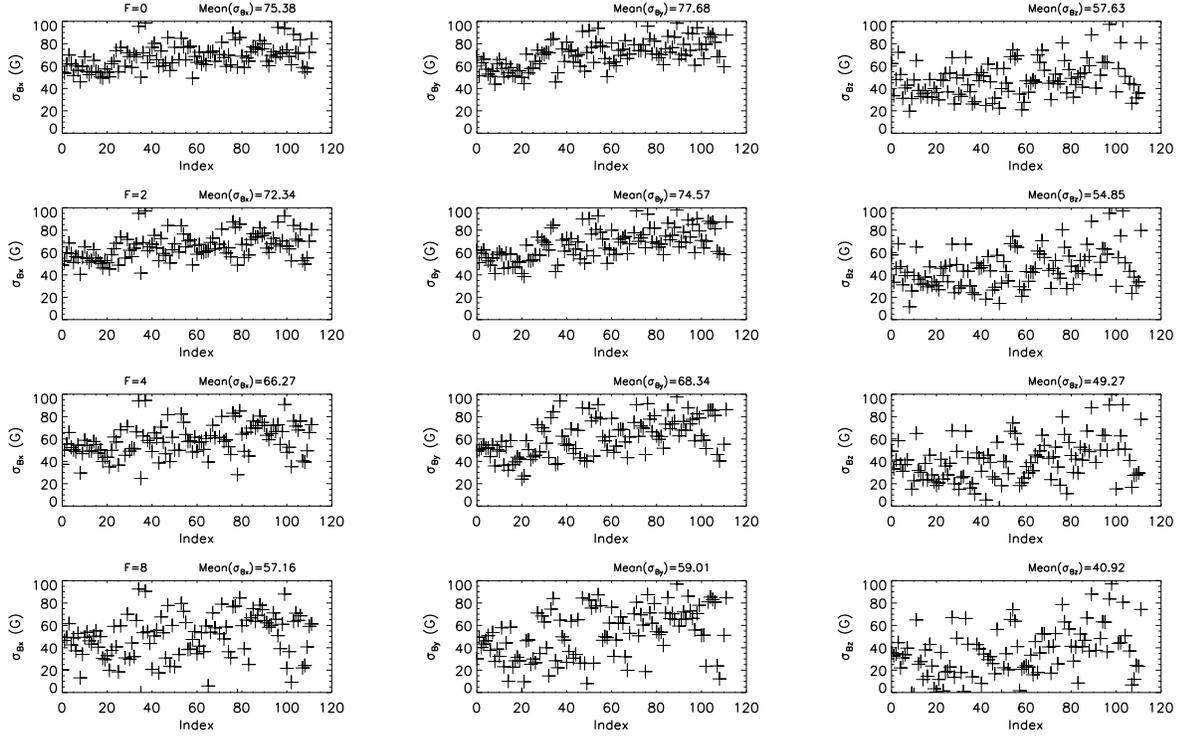}}

   \caption{The distribution of $\sigma$s of $B_{x}$, $B_{y}$, $B_{z}$ values for all active regions, and the values of $\sigma$s for active region with
   reduction resolutions are presented, the values of x-axis present the index of active region. Here the original resolutions of observations reduced by three different factors of 2, 4 and 8 are
   labeled by F=2, 4 and 8, respectively. } \label{Fig1-add-2}
\end{figure}

\begin{figure}

%%%I:\document-1\force-free-SP\readspbxbybz-dde-jpg-1sam-ps-art
   \centerline{\includegraphics[width=1\textwidth,clip=]{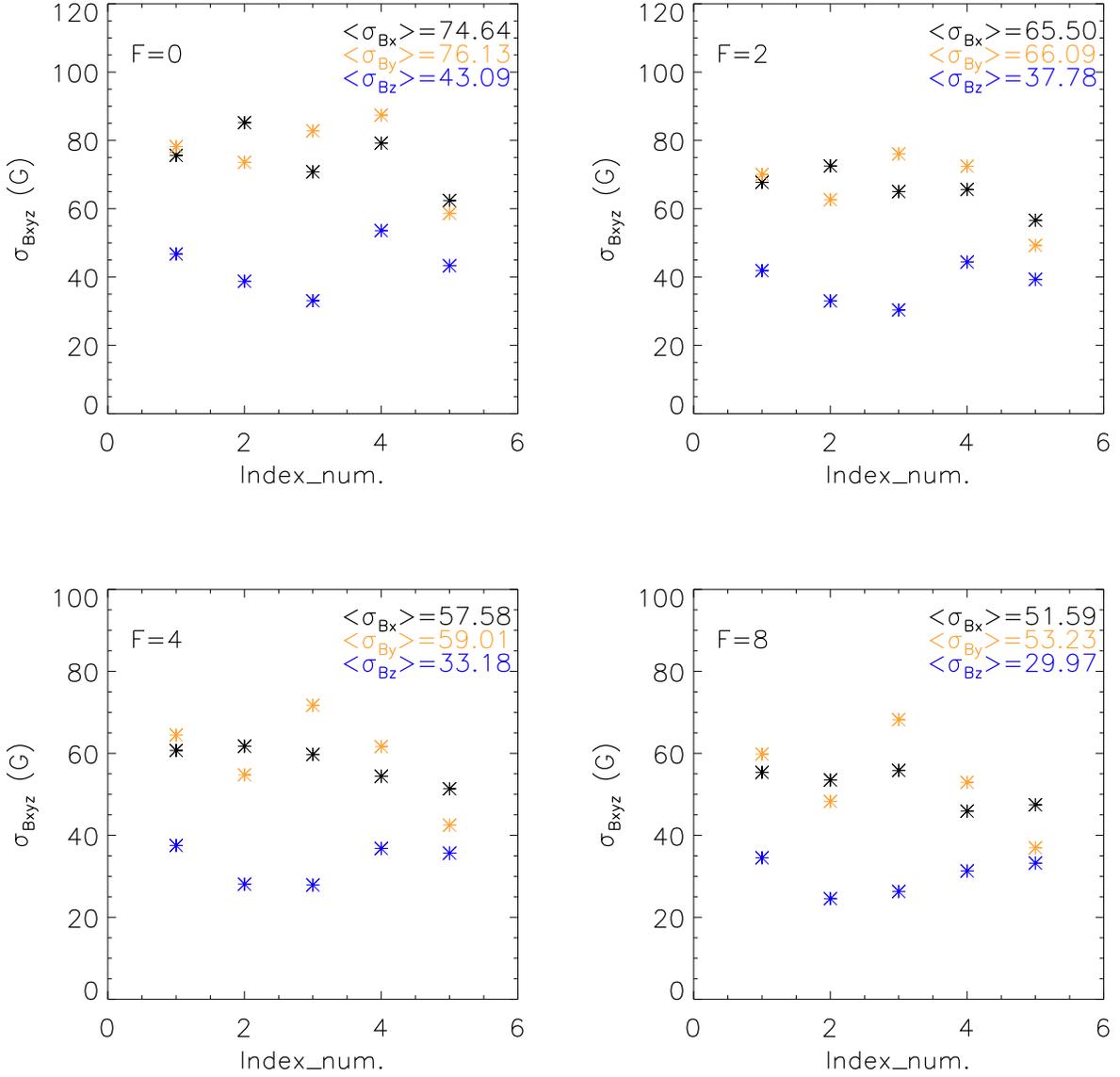}}

   \caption{The changes of $\sigma$s of $B_{x}$, $B_{y}$, $B_{z}$ with the reduction resolutions using Fig\ref{Fig1-add-1} as an example.
    Here the black, yellow and blue dots indicate $\sigma$s of $B_{x}$, $B_{y}$, $B_{z}$,
    five dots in each panel correspond five quiet regions labeled in Fig\ref{Fig1-add-1}, the values of x-axis correspond the indexs of five quiet regions in Fig\ref{Fig1-add-1},
    and the the original resolutions of observations reduced by three different factors of 2, 4 and 8 are
   labeled by F=2, 4 and 8, respectively. } \label{Fig1-add-3}
\end{figure}

\begin{figure}

%%%I:\SP\\readsav_plotzhif-sandian-sigma0123-sm248
   \centerline{\includegraphics[width=1\textwidth,clip=]{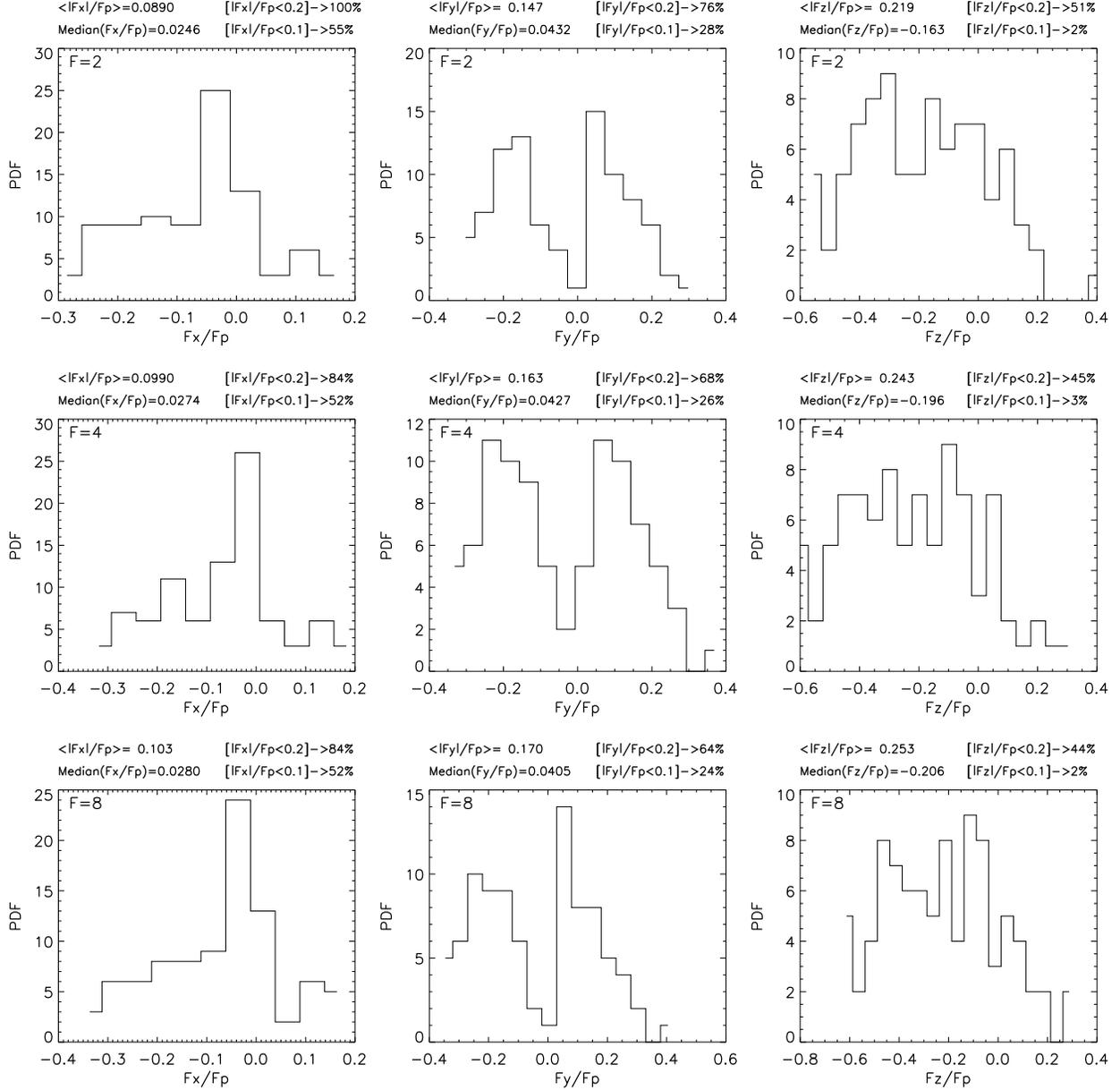}}

   \caption{PDF of $F_{x}/F_{p}$, $F_{y}/F_{p}$ and $F_{z}/F_{p}$ for the all
   selected magnetograms using 0 as threshold in calculations.
   The first row shows the results calculated from original resolution observations.
   The second, third and fourth rows shows the results computed from the reduce of original resolution by the factors of 2, 4 and 8.
   Also here, the mean values of absolute $F_{x}/F_{p}$, $F_{y}/F_{p}$ and
   $F_{z}/F_{p}$ are plotted and indicated by $<|F_{x}|/F_{p}>$,$<|F_{y}|/F_{p}>$, $<|F_{z}|/F_{p}>$, respectively.
   The median of $F_{x}/F_{p}$, $F_{y}/F_{p}$ and
   $F_{z}/F_{p}$ are plotted and indicated by $Median(F_{x}/F_{p})$,$Median(F_{y}/F_{p})$, $Median(F_{z}/F_{p})$, respectively.
   Additionally, the percentages of $F_{x}/F_{p}$, $F_{y}/F_{p}$ and $F_{z}/F_{p}$ less than 0.1 and 0.2 are given correspondingly.
   }\label{Fig2}
\end{figure}

To see the effects of observations resolutions on those parameters of $F_{x}/ F_{p}$, $F_{y}/ F_{p}$ and $F_{z}/ F_{p}$, the original resolutions of observations are reduced by three different factors of 2, 4 and 8, then the values of those parameters are calculated for different thresholds cases individually.
The thresholds maybe affected by reduction of resolutions at some extent, which are illuminated by Figures \ref{Fig1-add-2} for all active regions,
it gives the changes of $\sigma$s of $B_{x}$, $B_{y}$, $B_{z}$ with different resolutions, which are reduced by factors of 2, 4 and 8 (labeled by F=2, 4 and 8).
It can be found that the thresholds become smaller as the decreases of resolutions. Additionally, from its distributions it can be found that
 the amplitudes of decrease are
basically consistent for all active regions. At the same time Figure \ref{Fig1-add-3} gives the changes of $\sigma$s values with the variation of resolutions using Figure \ref{Fig1-add-1} as an example, here $\sigma$s of $B_{x}$, $B_{y}$, $B_{z}$ are indicated by the black, yellow and blue dots,
five dots in each panel correspond five quiet regions selected and labeled in Figure\ref{Fig1-add-1}, same as Figure \ref{Fig1-add-2} the resolutions reduced by three different factors of 2, 4 and 8, are labeled by F=2, 4 and 8, respectively. From Figure \ref{Fig1-add-2} and Figure \ref{Fig1-add-3}
it can be found that on the whole the amplitudes of changes with the reduction of resolutions are consistencies.
Figure \ref{Fig2} shows values of $F_{x}/ F_{p}$, $F_{y}/ F_{p}$ and $F_{z}/ F_{p}$ calculated for the case using 0 as threshold. Here the rows 1, 2 and 3 present the results based on the reduced original resolutions of observations by a factor of 2, 4 and 8. The same as Fig \ref{Fig1}, the averages, the median values and the percentages of two cases described same as Fig 1 are given in each plot. From Fig \ref{Fig2}, it can be found that $F_{x}/ F_{p}$, $F_{y}/ F_{p}$ and $F_{z}/ F_{p}$ become large as the decrease of resolutions on the whole, but the changes of these parameters are not very evident. It means that the resolutions effects on the values of $F_{x}/ F_{p}$, $F_{y}/ F_{p}$ and $F_{z}/ F_{p}$ is very faint when using 0 as the threshold.

\begin{figure}

%%%
   \centerline{\includegraphics[width=1\textwidth,clip=]{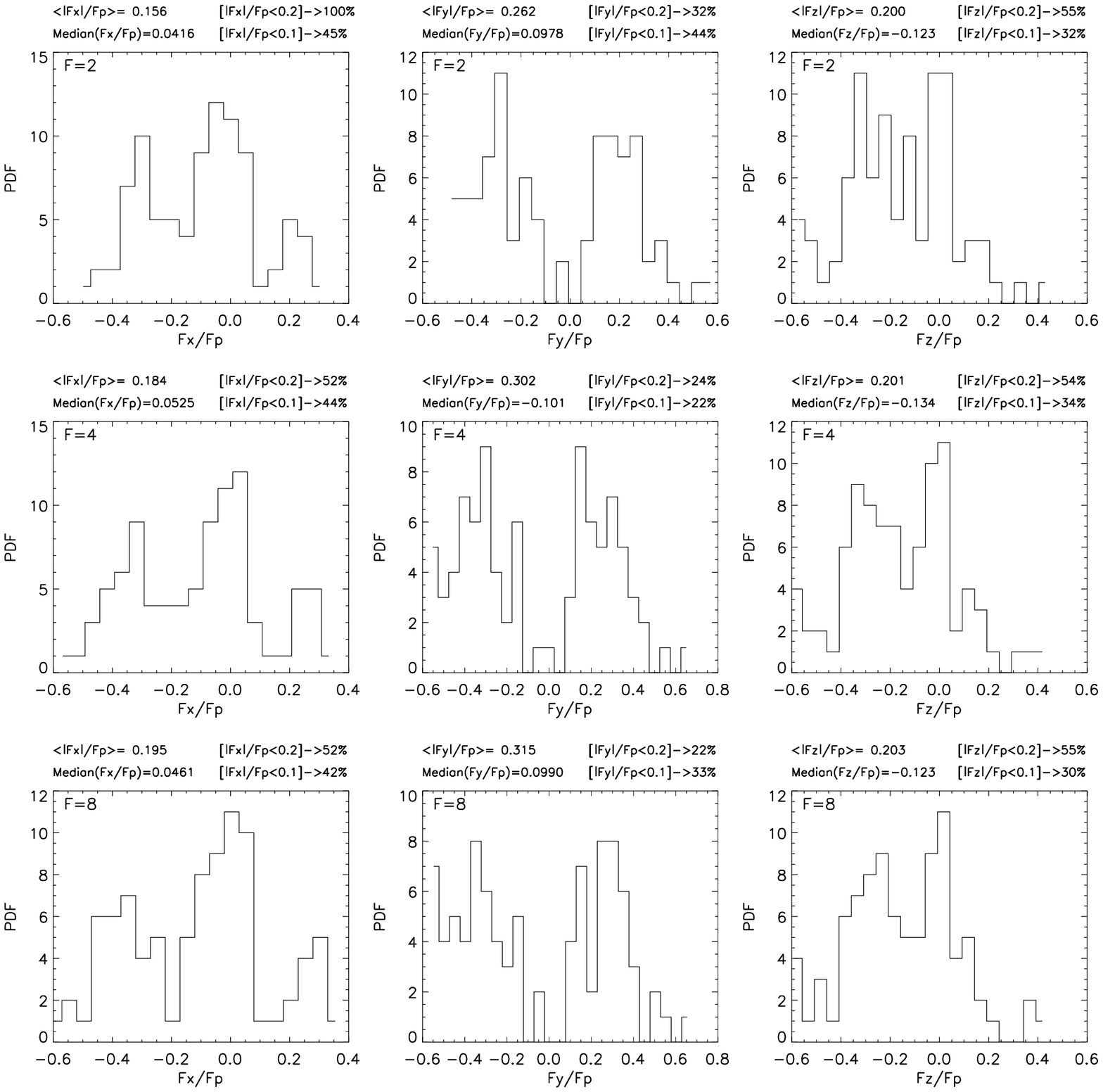}}

   \caption{The same as Fig2, but using 1$\sigma$ deviations of data noise as threshold.} \label{Fig3}
\end{figure}

\begin{figure}

%%%
   \centerline{\includegraphics[width=1\textwidth,clip=]{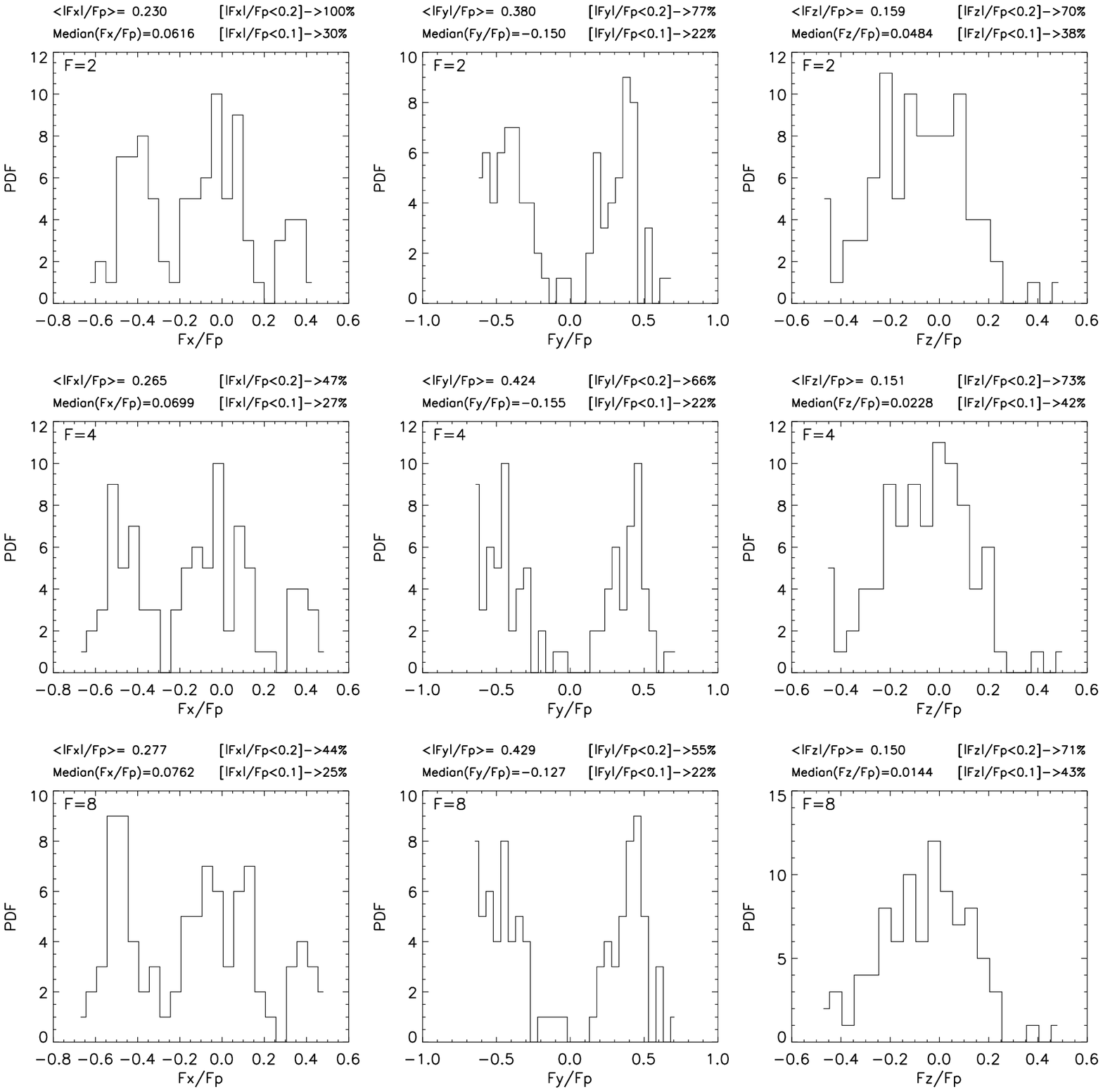}}

   \caption{The same as Fig2, but using 2$\sigma$ deviations of data noise as threshold.} \label{Fig4}
\end{figure}

\begin{figure}

%%%
   \centerline{\includegraphics[width=1\textwidth,clip=]{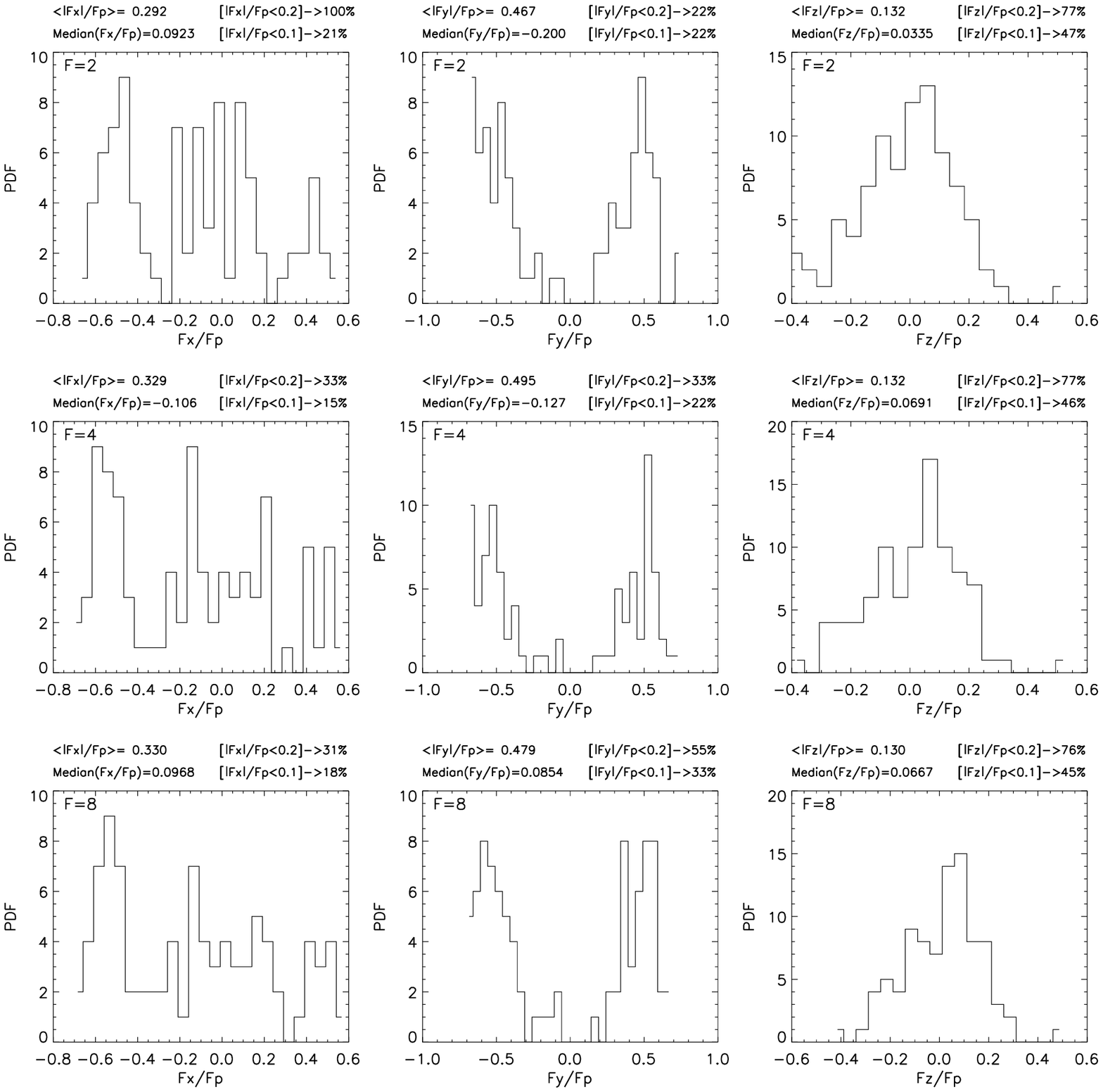}}

   \caption{The same as Fig2, but using 3$\sigma$ deviations of data noise as threshold. } \label{Fig5}
\end{figure}

The same as Fig \ref{Fig2}, Fig \ref{Fig3}, \ref{Fig4} and \ref{Fig5} show the distributions of $F_{x}/ F_{p}$, $F_{y}/ F_{p}$ and $F_{z}/ F_{p}$ using 1$\sigma$, 2$\sigma$ and 3$\sigma$ as thresholds, respectively. For 1$\sigma$, the results show the similar trend as using 0 as threshold, but the changes of $F_{z}/ F_{p}$
become week. For 2$\sigma$ and 3$\sigma$, on the whole it can be found that the values of $F_{x}/ F_{p}$, $F_{y}/ F_{p}$ become large as the resolutions decrease, while the values of $F_{z}/ F_{p}$ become small when the resolutions decrease.

From Fig \ref{Fig2}, Fig \ref{Fig3}, \ref{Fig4} and \ref{Fig5}, it can be found that the threshold used
and grid resolution have effects on the results at some extent. So in Fig \ref{Fig6} and \ref{Fig7} the average values and
standard deviation of $|F_{x}|/ F_{p}$, $|F_{y}|/ F_{p}$ and $|F_{z}|/ F_{p}$ vs the thresholds
and grid resolutions are given. From this two figures, it can be found that the amplitude of $|F_{x}|/ F_{p}$ and $|F_{y}|/ F_{p}$ increase as the
the increases/decreases of thresholds/resolutions. As for $|F_{z}|/ F_{p}$, the situations become more complex:
1. When the small thresholds used (threshold=0 and
threshold=1$\sigma$)
the amplitudes of $|F_{z}|/ F_{p}$ increase with the decrase of resolutions, while for the high thresholds (threshold=2$\sigma$ and threshold=3$\sigma$) they reverse;
 2. On the whole, the amplitudes of $|F_{z}|/ F_{p}$ decrease with the increases of thresholds used.

%To compare the changes amplitude of these parameters resulted from the effects of resolutions for different thresholds more clearly, the differences, which exist between original resolutions and the resolutions reduce by a factor of 8 for the above two cases ($F_{x}/ F_{p}$, $F_{y}/ F_{p}$ and $F_{z}/ F_{p}$ are less than 0.1 and 0.2), are given in table \ref{tbl-1} for each threshold. From table \ref{tbl-1}, it can be seen that on the whole the values of $F_{x}/ F_{p}$ and $F_{y}/ F_{p}$ tend to become large when reduce resolutions, while for $F_{z}/ F_{p}$ they become small after decrease resolutions. It can also be found that the changes amplitudes of the cases of $F_{(i=x,y,z)}/ F_{p}<0.2$ are more evident than those of $F_{(i=x,y,z)}/ F_{p}<0.1$ cases on the whole. Although there exist changes of these parameters of $F_{(i=x,y,z)}/ F_{p}$ for different resolutions, the distributions profiles of these parameters do not change evidently after the reduction of resolutions for each threshold.

To see the relations between $|F_{x}|/ F_{p}$, $|F_{y}|/ F_{p}$ and $|F_{z}|/ F_{p}$ and the magnetic
field strength, Fig \ref{Fig8} shows $|F_{x}|/ F_{p}$, $|F_{y}|/ F_{p}$ and $|F_{z}|/ F_{p}$ versus
the magnetic components of $B_{x}$, $B_{y}$, $B_{z}$, the transverse field ($B_{t}$=$\sqrt{B_{x}^{2}+B_{y}^{2}}$), total field strength $B$, magnetic azimuths ($\phi$=atan($B_{y}$/$B_{x}$)) and  magnetic inclination angles ($\theta$=atan($B_{z}$/$B_{t}$)) for using 2$\sigma$ data deviations as the threshold, where $B_{x}$, $B_{y}$ and $B_{z}$ are the average of their absolute values and the correlation coefficients between two
physical quantities in each panel are given correspondingly. In Fig \ref{Fig8} the conclusions can be given as follows:
1. From the plot a it can be found that the amplitudes of $|F_{x}|/ F_{p}$ increase as $B_{x}$ increase; 2. From the plot b it can be found that the amplitudes of $|F_{y}|/ F_{p}$ decrease as $B_{y}$ increase; 3. From the plot c, f and i it can be found that the values of $|F_{z}|/ F_{p}$ increase as $B_{x}$, $B_{y}$ and $B_{t}$ increase, repectively; 4. From the plot p it can be found that the amplitudes of $|F_{x}|/ F_{p}$ decrease as magnetic azimuths ($\phi$) increase; 5. From the plot q it can be found that the amplitudes of $|F_{y}|/ F_{p}$ increase as magnetic azimuths ($\phi$) increase; 6. From the plot u it can be found that the values of $F_{z}/ F_{p}$ decrease evidently as magnetic inclination angles ($\theta$) increase. 7. There is no evident correlation between $|F_{x}|/ F_{p}$, $|F_{y}|/ F_{p}$ and $|F_{z}|/ F_{p}$ and magnetic component can be revealed in the other plot. When other thresholds used in calculation, the relations between those parameters and magnetic components have the same
trends.

\section{DISCUSSIONS AND CONCLUSIONS}

The force-free properties of photosphere magnetic field can give the force-free coronal field extrapolations important references in practice.
In the applications of coronal force-free field extrapolations, the magnetic fields of photosphere that probably deviate force-free properties more or less should be regarded as force-free field. Hence, the force-freeness of photosphere magnetic field is worth to be studied enough to make the coronal force-free
field models more  reasonable.

In this paper, the force-freeness of photosphere magnetic field is studied statistically using active regions magnetograms observed by SP/Hinode.
Three parameters of $F_{x}/ F_{p}$, $F_{y}/ F_{p}$ and $F_{z}/ F_{p}$ defined in the equations (2)-(5) are used to inspect the extent of photosphere magnetic field. If the amplitudes of $F_{x}/ F_{p}$, $F_{y}/ F_{p}$ and $F_{z}/ F_{p}$ less than 0.1 the magnetic field can regard as sufficiently
force-free according to \cite{mat95}, and this criterion is also used in this paper to research the force-freeness of photosphere magnetic field.
The previous results from \cite{moo02}, it give the mean amplitudes and median of $F_{x}/ F_{p}$ are 0.04 and 0.01, for $F_{y}/ F_{p}$ they are 0.07 and -0.03, for $F_{z}/ F_{p}$ they are -0.17 and 0.13.
The results of \cite{tiw12} give the mean amplitudes and median of $F_{x}/ F_{p}$ are 0.11 and 0.05, for $F_{y}/ F_{p}$ they are 0.09 and -0.09, for $F_{z}/ F_{p}$ they are 0.23 and -0.22.
In this paper, the mean amplitudes and median of $F_{x}/ F_{p}$ are 0.075 and 0.021, for $F_{y}/ F_{p}$ they are 0.126 and 0.051, for $F_{z}/ F_{p}$ they are 0.172 and -0.112, here the results are calculated using 1$\sigma$ data noise level as the threshold which case is the same as \cite{tiw12}.
It can be found that the amplitudes of $F_{z}/ F_{p}$ are greater than those of $F_{x}/ F_{p}$ and $F_{y}/ F_{p}$ on the whole, therefore the main attentions are paid to $F_{z}/ F_{p}$ to study the force-free extent of photosphere magnetic field in previous researches.

\begin{figure}

%%%vsresandgrid
   \centerline{\includegraphics[width=1\textwidth,clip=]{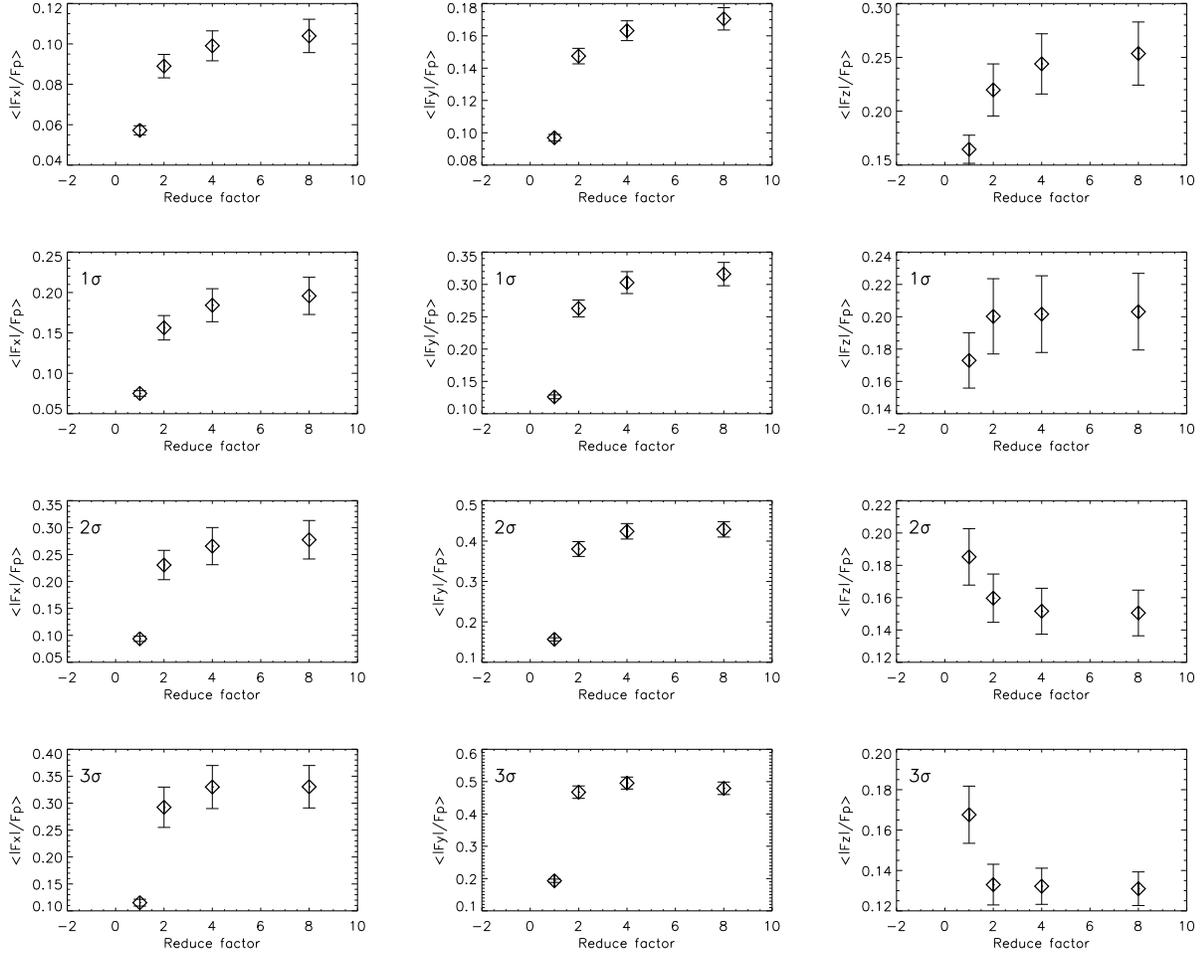}}

   \caption{The average values and
standard deviation of $F_{x}/ F_{p}$, $F_{y}/ F_{p}$ and $F_{z}/ F_{p}$ vs the grid resolution for the fixed threshold. The diamond and errors bar indicate
the averages and deviations, respectively. The row 1, 2, 3 and 4 indicate 0, 1$\sigma$, 3$\sigma$ and 3$\sigma$ thresholds used in calculations. X-axis is reduction factor done to original
resolution observations.} \label{Fig6}
\end{figure}

\begin{figure}

%%%vsresandgrid
   \centerline{\includegraphics[width=1\textwidth,clip=]{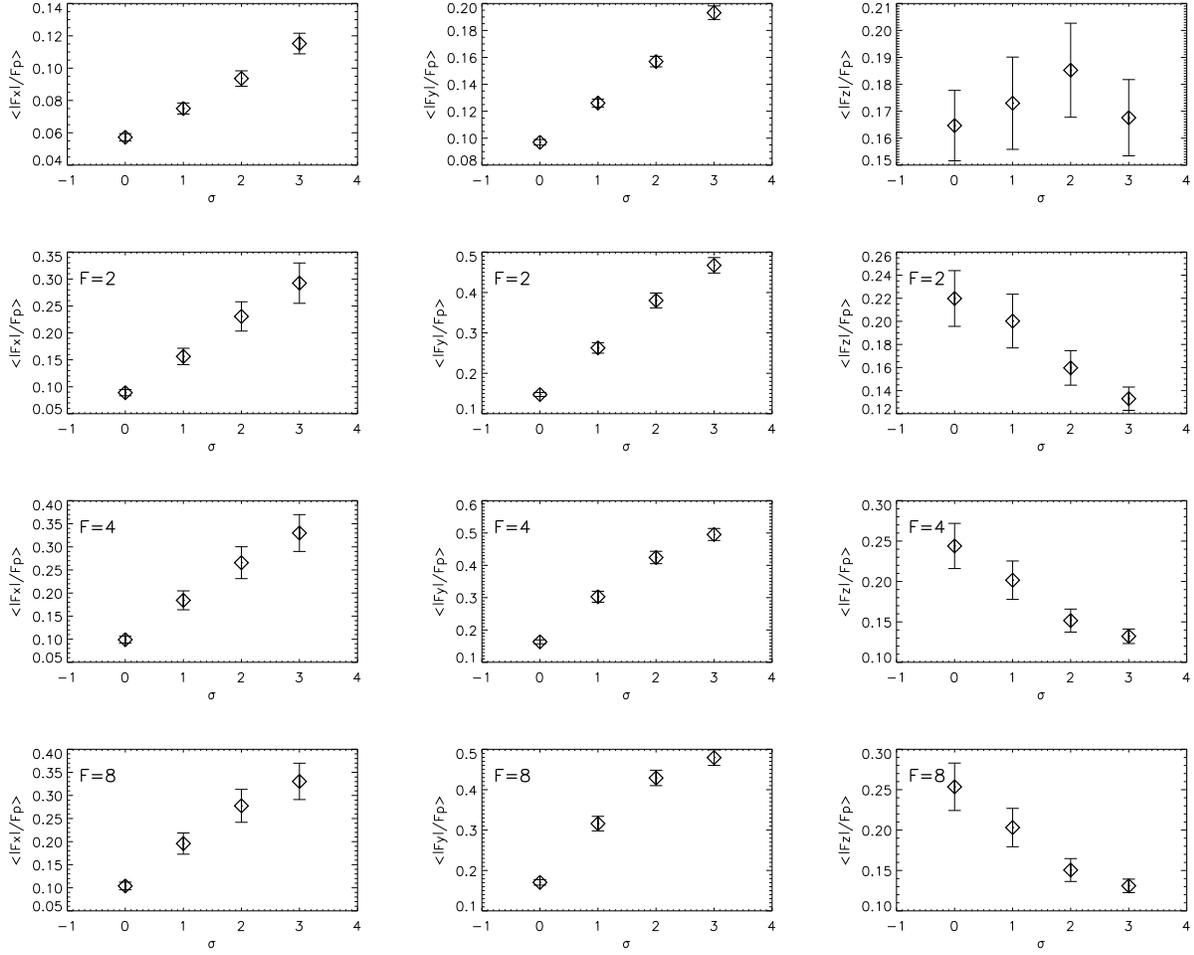}}

   \caption{The average values and
standard deviation of $|F_{x}|/ F_{p}$, $|F_{y}|/ F_{p}$ and $|F_{z}|/ F_{p}$ vs the threshold for the fixed resolution. The diamond and errors bar indicate
the averages and deviations, respectively. Here F=2, F=4 and F=8 indicate the reduction of original resolution by a factor of 2, 4 and 8, respectively.
X-axis mean different $\sigma$ selected.} \label{Fig7}
\end{figure}

The resolutions of magnetic field observations are always regarded as an important effect which may affect the integration values of those parameters, hence the results calculated from various resolutions are revealed, here the different resolutions are carried out by reducing the resolutions of SP observations. It is found that the amplitude of $F_{x}/ F_{p}$ and $F_{y}/ F_{p}$ become large as the resolutions decrease, while the amplitudes of $F_{z}/ F_{p}$ tend to be small as the resolutions decrease for the cases when the threshold are 2$\sigma$ and 3$\sigma$ data noise levels. The previous relative studies all based on the observations with the resolutions that is lower than those of SP observations, additionally, the amplitudes of $F_{x}/ F_{p}$ and $F_{y}/ F_{p}$ are all smaller than those of $F_{z}/ F_{p}$ in the previous studies. So at some extent the photosphere magnetic field maybe more force-free in the previous studies than in this study based on high resolutions observations.
For the case that there is no threshold used (namely, threshold=0), the amplitudes of of $F_{x}/ F_{p}$, $F_{y}/ F_{p}$ and $F_{z}/ F_{p}$ all become large, however the amplitudes of change are smaller than those of other thresholds used. It is noticed that it maybe not a reasonable data process when there is no threshold are used in data analysis. From the Fig \ref{Fig6}, it can be found that as the resolution increases the forces indicated by $F_{z}/ F_{p}$ can also increase,
it further illustrate the previous studies used low resolution data maybe more inclined to force-free at some extent.
Through calculations, it can be found that $F_{x}/ F_{p}$, $F_{y}/ F_{p}$ and $F_{z}/ F_{p}$ exhibited some complex situations when different thresholds and resolutions are selected. The reason maybe due to the changes amplitudes of different magnetic components ($B_{x}$, $B_{y}$ and $B_{z}$), through analysis it is found that the change trends of magnetic components is consistent, but there exist some small fluctuations for each magnetic component when difference thresholds and resolutions are
used, however there is no special regularity can be drawn.

\begin{figure}

%%%
   \centerline{\includegraphics[width=1\textwidth,clip=]{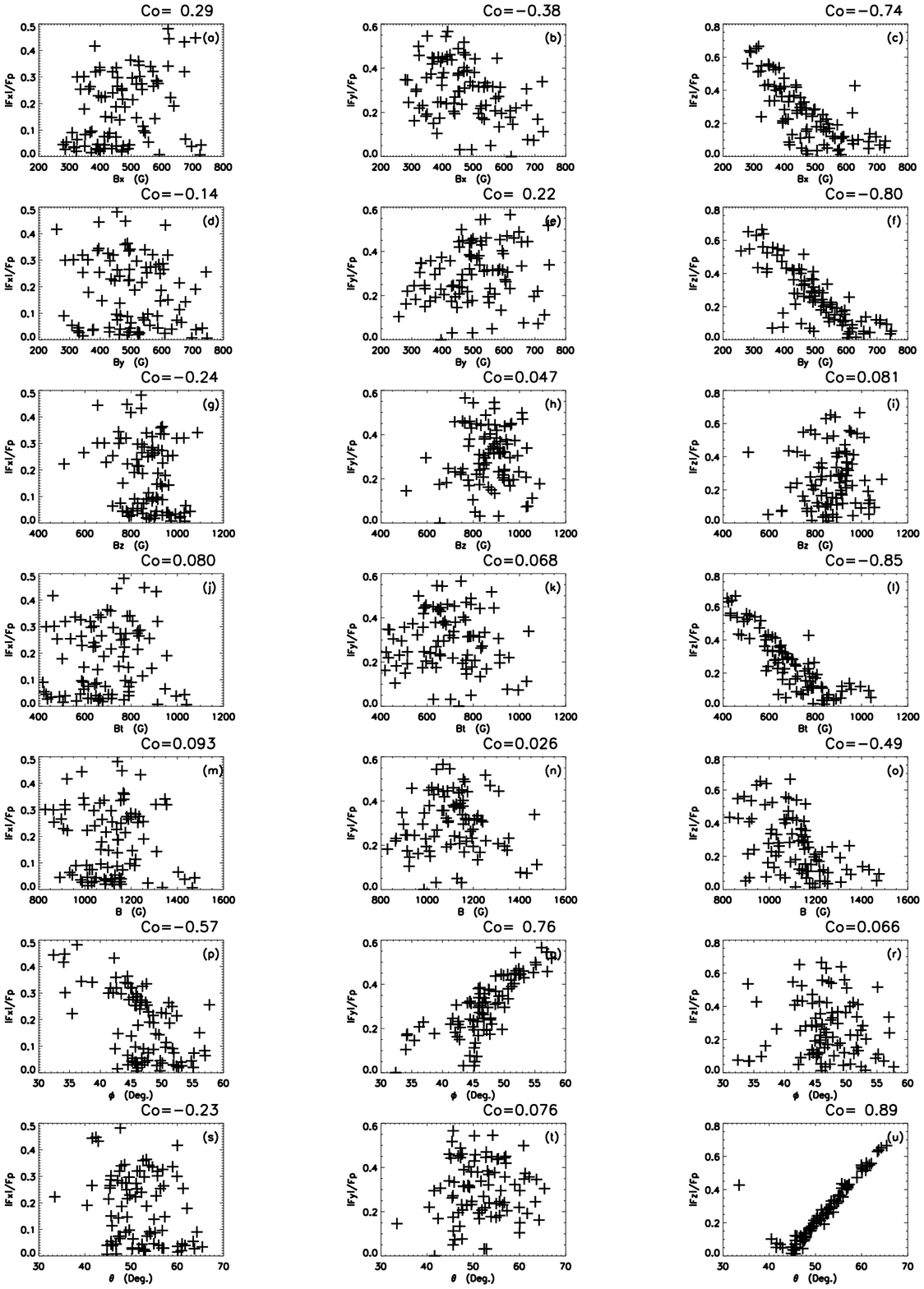}}

   \caption{The correlations of $|F_{x}|/F_{p}$, $|F_{y}|/F_{p}$ and $|F_{z}|/F_{p}$ Vs $B_{x}$, $B_{y}$, $B_{z}$, $B_{t}$, $B$, magnetic azimuths ($\phi$=atan($B_{y}$/$B_{x}$)) and  magnetic inclination angles ($\theta$=atan($B_{z}$/$B_{t}$)) calculated using 2$\sigma$ deviation as threshold.} \label{Fig8}
\end{figure}

The relationships between the parameters of $|F_{x}|/ F_{p}$, $|F_{y}|/ F_{p}$ and $|F_{z}|/ F_{p}$ and magnetic components are researched in this paper, here the magnetic components include $B_{x}$, $B_{y}$, $B_{z}$, $B_{t}$, $B$, azimuths $\phi$ and inclination angle $\theta$. On the whole it can be seen that the amplitudes of $|F_{x}|/F_{p}$ and $|F_{y}|/ F_{p}$ depend on  azimuths $\phi$ with the correlation coefficient of -0.57 and 0.76 ($\phi$ related to the relationships between $B_{x}$ and $B_{y}$ components), respectively, which are indicated in the plots of p and q in Fig \ref{Fig8}. It can be found that the amplitudes of $|F_{z}|/ F_{p}$ depend on the magnetic inclination angles evidently, which can be seen from the plot of u in Fig \ref{Fig8} with the correlation coefficient of 0.89.
It can also be found that$|F_{z}|/ F_{p}$ and $B_{x}$, $B_{y}$ and $B_{t}$ have some relations the correlation coefficient of -0.74, -0.80 and -0.85.
The relations between the amplitudes of $|F_{x}|/ F_{p}$, $|F_{y}|/ F_{p}$ and $|F_{z}|/ F_{p}$ and magnetic component of $B_{z}$ are very weak.
It should be noted that the magnetic components of $B_{x}$ and $B_{y}$ are the part of $B_{t}$, and high correlation coefficient (between $|F_{x}|/ F_{p}$, $|F_{y}|/ F_{p}$ and $|F_{z}|/ F_{p}$ and magnetic components) are more inclined to $B_{x}$ and $B_{y}$. Hence it reflect indirectly a fact in this data samples that the transverse field play more important roles than that of longitudinal field to determine the values of those parameters,
however it should keep in mind that the above rules (transverse field play important roles) are probably data dependent.

\acknowledgments
%The authors thank the anonymous referee for helpful comments and suggestions.
This work was partly supported by the Grants: 2011CB811401, KLCX2-YW-T04, KJCX2-EW-T07,
11203036, 11221063, 11178005, 11003025, 11103037, 11103038, 10673016, 10778723 and 11178016, the
Key Laboratory of Solar Activity National Astronomical Observations, Chinese Academy
of Sciences.


%% Here we use \plottwo to present two versions of the same figure,
%% one in black and white for print the other in RGB color
%% for online presentation. Note that the caption indicates
%% that a color version of the figure will be available online.
%%


\begin{thebibliography}{}

\bibitem[Aly(1984)]{aly84}
Aly, J.J.: 1984, \apj{} \textbf{283}, 349.

\bibitem[Aly(1989)]{aly89}
Aly, J.J.:
1989, \solphys{} \textbf{120}, 19.

\bibitem[Amari et al.(1997)]{ama97}
Amari, T., Aly, J.J., Luciani, J.F., Boulmezaoud, T.Z., Mikic, Z.:
1997, \solphys{} \textbf{174}, 129.

\bibitem[Gary \& Hurford(1994)]{gar94}
Gary, D. E. \& Hurford, G. J.:
1994, \apj{} \textbf{420}, 903

\bibitem[Hao \& Zhang(2011)]{hao11}
Hao, J., Zhang, M.: 2011, \apj{} \textbf{733}, L27

\bibitem[He \& Wang(2008)]{he08}
He, H., Wang, H.: 2008, \jgr{} \textbf{113}, A05S90

\bibitem[Jiang et al.(2013)]{jia13}
Jiang, C.W., Feng, X.SH.:
2013, \solphys{} \textbf{289}, 63%-77.

\bibitem[Lites et al.(1999)]{lit99}
Lites, B. W., Rutten, R. J. \& Berger, T. E.:
1999, \apj{}, \textbf{517}, 1013.

\bibitem[Lin et al.(2004)]{lin04}
Lin, H., Kuhn, J.R., Coulter, R.:

\bibitem[Liu et al.(2013)]{liu13}
Liu, S., Su, J.T., Zhang, H.Q., Deng, Y.Y., Gao, Y., Yang, X., Mao, X.J.:
2013, $Publications of the Astronomical Society of Australia$,  \textbf{30}, e005

\bibitem[Liu et al.(2011a)]{liu11a}
Liu,S., Zhang, H.Q. \& Su, J.T.:
2011, \solphys{} \textbf{270}, 89%-481.

\bibitem[Liu et al.(2011b)]{liu11b}
Liu,S., Zhang, H.Q., Su, J.T., Song, M.T.:
2011, \solphys{} \textbf{269}, 41

\bibitem[Low(1985)]{low85}
Low, B.C.:
1984, in Measurements of Solar Vector Magnetic Fields, Vol. 2374,
ed. M. J. Hagyard (Washington, DC: NASA), 49

\bibitem[Metcalf et al.(1995)]{mat95}
Metcalf, T. R., Jiao, L., McClymont, A. N., Canfield, R. C.,
Uitenbroek, H.:
1995, \apj{}, \textbf{439}, 474

\bibitem[Metcalf et al.(2006)]{mat06}	
Metcalf, T. R., Leka, K. D., Barnes, G., Lites, B. W., Georgoulis, M. K., Pevtsov, A. A.,
Balasubramaniam, K. S., Gary, G. A., Jing, J., Li, J.:
2006, Solar Phys{} \textbf{237}, 267

\bibitem[Moon et al.(2002)]{moo02}	
Moon, Y., Choe, G. S., Yun, H. S., Park, Y. D., Mickey, D. L.:
1995, \apj{} \textbf{568}, 422.

\bibitem[Molodenskii (1967)]{mol67}	
Molodenskii, M.M.:
1967, Soviet Astronomy, \textbf{12}, 585.

\bibitem[Mikic et al.(1994)]{mic94}
Mikic, Z.; McClymont, A. N.: 1994, in Solar Active Region Evolution:
Comparing Models with Observations, Vol68. ASP Conf. Ser., p.225.


\bibitem[Kosugi et al.(2007)]{kos07}
Kosugi, T., Matsuzaki, K., Sakao, T., Shimizu, T., Sone, Y.,
Tachikawa, S., Hashimoto, T., Minesugi, K., Ohnishi, A., Yamada, T.:
2007,  \solphys{} \textbf{243}, 3.

\bibitem[Sakurai (1981)]{sak81}
Sakurai, T.: 1981, \solphys{} \textbf{69}, 343.

\bibitem[Song et al.(2006)]{son06}
Song, M.T., Fang, C., Tang, Y.H., Wu, S.T., Zhang, Y.A.:
2006, \apj{} \textbf{649}, 1084.

\bibitem[Tiwari(2012)]{tiw12}
Tiwari, S. K.:
2012, \apj{} \textbf{744}, 65.

\bibitem[Wang(1997)]{wan97}
Wang, H.:
1997,  \solphys{} \textbf{174}, 265.


\bibitem[Wang et al.(1994)]{wan94}
Wang, T., Zhang, H. and Xu, A.:
1994,  \solphys{} \textbf{155}, 99.%99-112

\bibitem[Wheatland et al.(2000)]{whe00}
Wheatland, M.S., Sturrock, P.A., Roumeliotis, G.: 2000, \apj{}
\textbf{540}, 1150.

\bibitem[Wiegelmann (2004)]{wie04}
Wiegelmann, T.: 2004, \solphys{} \textbf{219}, 87.

\bibitem[Wiegelmann et al.(2012)]{wie12}
Wiegelmann, T.,  Sakurai, T.:
2012, Living Reviews in Solar Physics, vol. 9, no. 5


\bibitem[Wiegelmann et al.(2006)]{wie06}
Wiegelmann, T.; Inhester, B.; Sakurai, T.:
2006, \solphys{} \textbf{233}, 215%-232.

\bibitem[Wu et al.(1990)]{wu90}
Wu, S.T., Sun, M.T., Chang, H.M., Hagyard, M.J., Gary, G.A.: 1990,
\apj{} \textbf{362}, 698.

\bibitem[Yan \& Sakurai(2000)]{yan00}
Yan, Y., Sakurai, T.:
2000, \solphys{} \textbf{195}, 89.


\end{thebibliography}
\end{document}